\documentclass[aps,pra, onecolumn,14 pts,a4paper]{revtex4}
 \pdfoutput=1

\usepackage{amssymb,amsmath,amsfonts,graphicx}
\usepackage{bm}
\usepackage{epsfig,color,times}
 \usepackage{bm,color}
\usepackage{graphicx}
\usepackage{amssymb}
\usepackage{epstopdf}

\begin{document}

\title{Leray turbulence: what can we learn from acceleration compared to velocity}

\author{Yves Pomeau$^1$ and Martine Le Berre$^2$}
\affiliation{ $^1$ Ladhyx (CNRS UMR 7646), Ecole Polytechnique, 91128 Palaiseau, France 
\\$^2$  Ismo (CNRS UMR 8214), Université de Paris-Sud,
91405 Orsay, France}

\date{\today }

\begin{abstract}
In a recent paper we presented evidence for the occurence of Leray-like singularities with positive Sedov-Taylor exponent $\alpha$  in turbulent flows  recorded in Modane's  wind tunnel,  by  looking at simultaneous  acceleration and velocity records. Here we  use another tool which  allows to get other informations on the  dynamics of turbulent bursts. We compare  the structure functions for velocity and acceleration in  the same turbulent flows.  This shows the possible contribution of other types of self-similar solutions  because  this new study shows that statistics is seemingly dominated by singularities  with small positive or even negative values of the exponent $\alpha$, that corresponds to ''weakly singular'' solutions with  singular  acceleration, and regular velocity. 
We present several reasons explaining that the  exponent $\alpha$ derived from the  structure functions curves, may look to be negative.
\end{abstract}

\maketitle

   \section{Introduction}
    \label{Introduction} 

     Many natural and man-made flows have typically large to very large Reynolds number and negligible Mach number, the limit we shall deal with. The neglect of viscosity and compressibility leads  to the Euler equations  \cite{Euler}, 
    \begin{equation}
{\partial_{t}{\bf{u}}} + {\bf{u}}\cdot \nabla {\bf{u}} = - \nabla p
\textrm{,}
\label{eq:Euler1}
\end{equation}
and 
\begin{equation}
\nabla \cdot {\bf{u}} = 0
\textrm{,}
\label{eq:Euler2.0}
\end{equation}
where $\partial_{t}$ is for the time derivative, the vector  ${\bf{u}}({\bf{r}}, t)$ is the local value of the fluid velocity at time $t$, boldface being for vectors, and $p$ is the pressure, a gauge function allowing to satisfy the condition of incompressibllity (\ref{eq:Euler2.0}). The nabla sign is for the gradient with respect to coordinate ${\bf{r}}$ and the mass density has been set to $1$.  

The next question is: what is predicted by Euler equations? 
 We do not know yet if smooth and bounded velocity fields in three space dimensions always yield a smooth solution at any later time. 

In 1934 Leray \cite{leray} suggested to explain the irregularity of turbulent flows by a loss of predictability linked to the occurrence of singularities in the solutions of this evolution equations. Such singularities would forbid to continue the solution afterwards within the framework of deterministic fluid equations. Leray went further than that and wrote down the equation for self-similar solutions of the Navier-Stokes equation (including the viscosity - see below) describing the evolution from smooth initial data to a solution singular at one point of space and time (his equation (3.11)).  

We have  recently presented  evidence for the occurrence of Leray-like singularities in a high-speed wind tunnel, as recorded by a hot wire\cite{PC}. Such singularities manifest themselves by the property that  {\textit{large}} values of the velocity fluctuations $u$  are correlated to large values  of  the acceleration $a$ at the same point and same time,  those large fluctuations being linked by a relation of the form
    \begin{equation}
a \sim u^{z} \qquad z\approx3
\textrm{.}
\label{eq:a-u3}
\end{equation}
On the contrary Kolmogorov (1941) scaling law yields $z=-1$,  and so predicts a vanishing acceleration when the velocity is large and conversely a large acceleration and a small velocity at the same point of space-time in the fluid, exactly the opposite of what is observed. From this observation we inferred  that the data are quantitatively well explained by the occurrence of Leray-like singularities of Euler fluid equations with exponents derived by assuming scaling laws coherent with the local conservation of energy, a point detailed just below.

As did Leray \cite{leray} let us suppose there is a singularity in finite time by the evolution of the flow velocity and that this singularity is of the self-similar type. The corresponding solution of the Euler equations is of the type 
 \begin{equation}
 {\bf{u}}( {\bf{r}}, t) = (t^*- t)^{-\alpha}  {\bf{U}} ( {\bf{r}}(t^*- t)^{-\beta})\textrm{,}
\label{eq:self-sim}
\end{equation}
  where $t^*$ is the time of the singularity (set to zero afterwards), where $\alpha$ and $\beta$ are exponents to be found and where  $ {\bf{U}}(.)$ is to be derived by solving Euler equations. Such a velocity field is a solution of Euler equations if 
   \begin{equation}
     \alpha + \beta=1
 \textrm{.}
\label{eq:sumexp}
\end{equation}    
 
In \cite{PC} our analysis of the experiments performed in Modane pointed to values of $\alpha$ and $\beta$ both close to $1/2$.  We argued that those values could be associated to two (different) conservation laws, each one corresponding to a set of exponents,  one is $\alpha=\beta=1/2$  which ensures the conservation of the circulation around the the singularity, the other one  providing the conservation of energy in the singular domain is  $\alpha = 3/5$ and $\beta = 2/5$,  which is related to the Sedov-Taylor scaling. The former  set leads to $z=3$,  the latter, which   leads to $z=8/3$, in slightly better agreement with the data than $z = 3$.

\begin{figure}
\centerline{ 
(a)\includegraphics[height=1.75in]{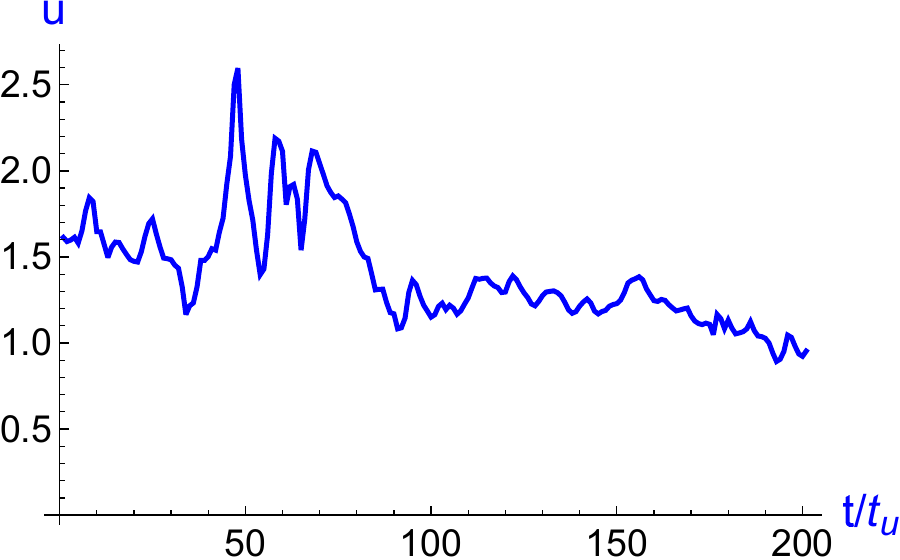}
  }
  \centerline{ 
(b)\includegraphics[height=1.5in]{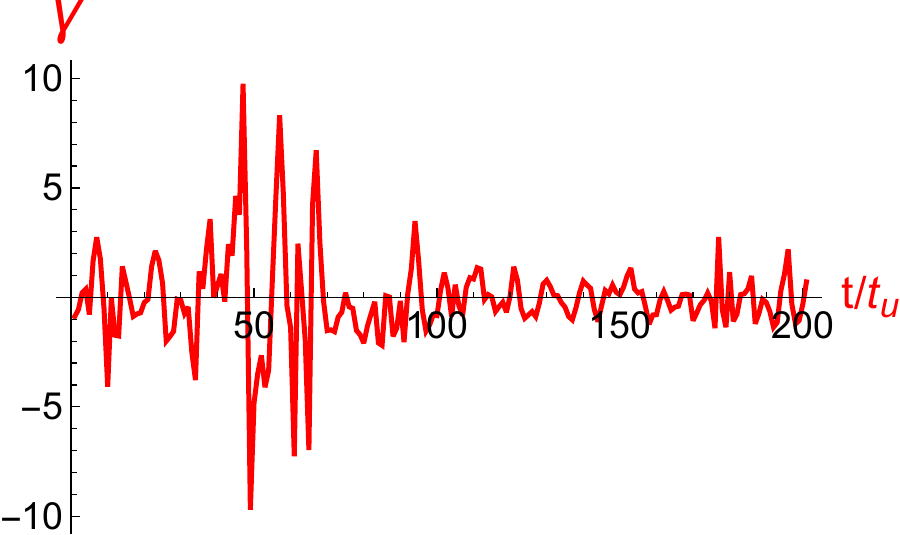}
  }
\caption{ Example of turbulent burst  in  Modane wind tunnel experiment performed by Y. Gagne et al. in the $1990$'s, with  $t_{u}= 1/25ms$ and other parameters given in \cite{PC}. The velocity fluctuation $u$ in (a) and the acceleration $\gamma$ in (b) are scaled to their respective r.m.s.   $\gamma$ is typically $5$ times larger than $u$ in turbulent bursts.  }
\label{fig:pic-mod1}
\end{figure}

Here we look at other possible values for the  exponents $(\alpha, \beta)$. At this time it is unknown if an unique value of the exponents exists, as resulting from the equation of self-similarity or if more than a single value exists. Looking carefully to the experimental data, it  appears that the divergence of the velocity, if there is any, is fairly weak.  At first sight comparing the amplitude of the velocity  fluctuations and  of the acceleration versus time, we observe that the former are noticeably smaller  than the latter (in units of their respective r.m.s), as illustrated in Fig.\ref{fig:pic-mod1}  showing  the temporal traces of acceleration and velocity in the case of a medium intensity burst. Note that  the existence of several values for the  exponents $(\alpha, \beta)$   is  all the same compatible with the relation (\ref{eq:a-u3}) found to agree very well with the experimental data of Modane,
because  the analysis done in  ref \cite{PC} correlates large values of the acceleration and large values of the velocity at the same point in space-time. It could well be that (\ref{eq:a-u3}) does not describe {\it{all}} fluctuations of the velocity for a given value of the acceleration that contribute to averages so that   the singularities considered in  \cite{PC}  are only part of a larger set of singular solutions, with different values of the exponent $\alpha$, in the range $\alpha +1> 0$. The latter condition  is necessary to yield a singularity of the acceleration, as observed,
 because  the exponent for the time dependence of the acceleration is $(- (\alpha + 1))$ and has to be positive.
Therefore we argue that it is even possible that the exponent $\alpha$ is negative for some solutions of the Euler-Leray equation (\ref{eq:Euler1sm}).   Actually  we show that a subset of singular solutions with negative $\alpha$ values 
 \begin{equation}
 -1< \alpha < 0
\label{eq:alphaneg}
\end{equation}
 is  consistent with the observation that the structure function built with the velocity does not seem to become singular at short distances, whereas the structure function built with the acceleration becomes more and more peaked near $r = 0$ as the exponent $n$ increases. 
  This result allows to define  such "weak" singularities as singularities of the acceleration but not of the velocity. They are obviously unseen when one focus on the correlation of the largest values of the observed acceleration and velocities,
  while the  present study comparing the structure functions for velocity and acceleration implies that statistics is dominated by events with small positive or even slightly negative values of $\alpha$. 
  
 It could also be that even in presence of solutions with $\alpha$ positive, the statistical analysis performed here select another family of solutions than in\cite{PC}, a subtle point related to the dilation invariance of Euler equation which have a family of solutions, $\mu^{-1} u(\mu x, t)$,  parametrized by  their amplitude $\mu^{-1}$, as discussed in Sec. \ref{dilation}. Below  we point out that close to a singular point, a self-similar solution with  given ($\alpha, \mu$)  values behaves differently than the one with ($\alpha, 1$), so that an experiment sensitive to the transient  behavior towards the singular time may lead the observer to conclude that  solutions with negative $\alpha$ values exist in the flow although it is false. We show that such misleading effect   could result from the existence of a large proportion of  weak amplitude singularities as compared to large amplitude ones. 
 
   There is another point related to what we call "singular event". Such an event is seen as a large fluctuation in the time records of velocity  but such records deal with fluid mechanics of real fluids. It means two things. First viscosity should play a role, likely (but not necessarily, see below) by smoothing the singularity when its typical size becomes too small. Because this happens when the velocity and acceleration become very large, the net effect of the viscosity should be to quell the growth of the velocity and of the acceleration. This is in fair agreement with the fact that the short distance behavior of the structure function is less divergent than predicted by the inviscid Euler-Leray equations. Another property of the real data is that after the singularity takes place a decay phase, see the example shown in  Fig.\ref{fig:pic-mod1},  that is, at least partly recorded as a large fluctuation.  This "post singularity" dynamics depends on the way the system crosses the singularity, which relies in particular on dissipative effects induced either by viscosity or compressibility and emission of sound. How such post singularity events contribute to the structure function is unknown, but it is reasonable to assume that they do it with exponents smaller than the pre-singularity part just because this relies on dissipation allowing to cross the inviscid singularity time. 
  
 With the variable ${\bf{R}} =  {\bf{r}}(- t)^{-\alpha}$,  the Euler equations become a set of equations (the Euler-Leray equations) for $ {\bf{U}}({\bf{R}})$. The incompressibility condition is 
\begin{equation}
\nabla \cdot {\bf{U}} = 0
\textrm{.}
\label{eq:Euler2ss}
\end{equation}
There exists also an extended version of the similarity equation with an explicit time dependent part. This is found by using as variable $\tau = - \ln(t^* - t)$ and by keeping a velocity field depending on  ${\bf{r}}$ via the stretched radius   ${\bf{R}} $. This yields the modified Euler-Leray equation  
 \begin{equation}
 \frac{\partial {\bf{U}}}{\partial \tau} + \alpha{\bf{U}}  + \beta {\bf{R}}\cdot \nabla {\bf{U}} + {\bf{U}}\cdot \nabla {\bf{U}} + \nabla P = 0
\textrm{.}
\label{eq:Euler1sm}
 \end{equation}
The present note aims at giving support to the link between the occurrence of singularities and the observation of strong gradients. This relies on a precise analysis of experimental records of velocity fluctuations in Modane's wind tunnel. Instead of conditioning the measured values of the velocity to the one of acceleration, as done in \cite{PC},  we introduce in next section a statistical analysis taking into account the random occurrence in space and time of Leray singularities. Assuming that self-similar solutions are  mainly responsible for the striking divergence of high-order structure functions at small spatial scales, we derive an estimation of their power-law dependence on the form $r^{p(n)}$. We also  indicate how the space-time density of singular events can be expressed in terms of the parameters of turbulence. This problem is discussed at the end of this subsection. In subsection \ref{exp} we present the real structure functions deduced from the data taken in Modane's wind tunnel, namely  the same data as in \cite{PC}. The  striking results are discussed and interpreted in Sec.\ref{discussion} taking into account  the role of the viscosity in the ultimate stage of singularity formation (a  question  studied in more details in \cite{PC}), the role of the dilation invariance on the measurements, and other possible causes.

  \section{ Structure functions and large fluctuations}
    \label{large fluctuations} 
 
We look at the structure functions traditionally defined as an ensemble average  of the spatial increment ${\bf{\delta X}}$ of a vector field ${\bf{X}}(r,t)$ in between two points   $r_0$ and $(r + r_0)$,  for an arbitrary time, the same time. The ensemble average becomes a sliding average over $r_{0}$ when the turbulence is homogeneous, a condition assumed hereafter, then  structure functions only depend on the distance  $r$. One has
 \begin{equation}
{\mathcal{S}}_{n}(r)=< {\bf{\delta X}}^{n}>
\label{eq:Sn}
\end{equation}
where $\delta{X} ={\bf{X}}(r+r_{0}) - {\bf{X}}(r_{0})$. We consider  such moments associated to the velocity, namely for ${\bf{X}}={\bf{u}}$, and also to the Eulerian acceleration defined as ${\bf{X}}={\partial_{t}{\bf{u}}}$. 

For  large $r$, the structure functions tends to a constant, see Figs.\ref{fig:Sn},  given by the sum $\mathcal{S}_{n} \to \sum_{p=1} ^{n}C_{p}^{n} <X^{p}><X^{n-p}>$. In the following we focus on the behavior of ${\mathcal{S}}_{n}(r)$ at short distance for even $n$.
As we noticed in ref \cite{PC}, the velocity fluctuations are not very far from being Gaussian, which makes large fluctuations quite unlikely. On the contrary the probability distribution of the acceleration has much wider wings than a Gaussian, that gives them a fairly large probability of being big. Since we look for properties linked to singularities, and so to big fluctuations, we thought it better to look at the structure function built with the acceleration field instead of the velocity field. Whereas we found that the velocity fluctuations hardly show any trace of singular events in the $r$-dependence of the structure function,  we found that, on the contrary, the structure functions built with the acceleration show a clear, not to say obvious, evidence of singularities in the short range behavior of the structure function, see  Figs.\ref{fig:Sn}.  

\subsection{Contribution of quasi-singular solutions}  
\label{contribution}
 Let us  give first  an estimate of the contribution of quasi-singular events to structure functions. By such a quasi-singular event we mean a solution of the evolution equation (Euler equation in fluid mechanics of inviscid fluids) becoming singular in finite time at a given point of space and time. This is said to be quasi-singular because it cannot be a singularity in the exact mathematical sense in real life:  Euler equations have to break down at small scales, either because of the effect of viscosity if the Reynolds number in the Euler singularity decreases too much, or if this is not the case (no decrease of the Reynolds number), other physical effects like compressibility should become relevant. We comment  on those possibilities at the end of this section including our study based on experimental observations. 

In the following we drop the boldface notation for the vector field, and denote as $a(r, t)$ the field under consideration to remind that we shall focus on acceleration. Close to a singular event of type $q$ (see just below) occurring at  the space-time point $(r_i^{*}, t_i^{*})$, the solution of the fluid equations  is of the form $  a_s(r - r_i^{*},  t_i^{*} - t , q)$ where  $a_s(r, t, q)$ is singular at $r = t =0$ by definition, and $q$ is a set of parameters taking into account the possibility that there is more than one kind of singular solution, either because of a multiplicity of solutions like in the eigenvalues of a standard matrix or because of various symmetries of those solutions, either geometrical (by rotation) or because of the dilation symmetries of the Euler equations. 
For example in the case of a solution like (\ref{eq:self-sim}) with given exponents $\alpha_{i}, \beta_{i} $ and dilation parameter $\mu_{i}$, setting $q_{i}= \{\alpha_{i}, \beta_{i}, \mu_{i},\tilde{q}_{i}\}$, the acceleration close to a singular event of $\tilde{q}_{i}$-type near $(r_i^{*}, t_i^{*})$ can be written as
\begin{equation}
a_{s}(r, t, q_{i}) = \mu_{i}^{-1}\frac{1}{(t_{i}^{*}- t)^{\alpha_{i}+1} } A_{s} (\mu_{i} \frac{  r- r_{i}^{*}   }{  (t_{i}^{*}- t)^{\beta_{i} }}, \tilde{q}_{i}) 
\textrm{,}
\label{eq:a(q)}
\end{equation}

 Because the singularities are localized in space and time, one does not expect them to overlap. Instead one even expects that they tend to interact by repulsion because the growth of a singularity  exhausts the local possibility of  another one  growing nearby at about the same time, which induces a kind of repulsion between singularities \cite{CJ}. To streamline the explanations we shall assume that singularities do not interact at all so that the velocity field created by all singularities is a superposition of independent singular solutions with random choices of $t_i$ and $r_i$ with a given density $\nu(q_{i})$  in space and time. Correlated singularities should be taken into account, but we skip them because it would add lot of difficulties for  estimating $S_{n}(r)$ compared to the rough one presented now. 
 
We make the  hypothesis that  high order moments $S_{n}(r)$ ($n$ large)  are dominated  by the contribution of   singularities.  If the parameters labelled $q$ form a discrete set
$\{q_{i} \}$, the structure function can be approximated by a sum of $\delta a_{s} $ over  space-time domains $\mathcal{D}_{i}(q_{i}) $ labelled by the discrete index $i$,    
\begin{equation}
{\mathcal{S}}_n (r) \approx  \sum_{i}  \nu(q_{i})   \int_{\mathcal{D}_{i}(q_{i})}{\mathrm{d}} r_0 \; {\mathrm{d}} t  \;   ( a_{s}(r_{0}+r -  r_i^{*}, t - t_i^{*}, q_{i})  -  a_{s}(r_{0}  -  r_i^{*}, t - t_i^{*} ,  q_{i}) )^n 
\textrm{.}
\label{eq:sni}
\end{equation}

Going to a continuum of possible quasi-singularities,  let define $\nu(q)$  as the density of  quasi-singularity $q$ per unit time and unit volume. The mean number  of quasi-singularities per unit time and unit volume, is
\begin{equation}
 \tilde{\nu}=\int{\mathrm{d}}q  \; \nu(q) 
\textrm{,}
\label{eq:probaas4.1}
\end{equation}
which has the physical dimension of a frequency times an inverse volume. 
The sum (\ref{eq:sni}) 
becomes
\begin{equation}
\mathcal{S}_{n}(r) \approx  \int{\mathrm{d}}q  \; \nu(q)   \int {\mathrm{d}}r_{0}  \int{\mathrm{d}}t   \;  ( a(r +r_{0}, t \vert q)  - ( a(r_{0} , t \vert q) )^n  
\textrm{.}
\label{eq:sn2}
\end{equation}

Equations (\ref{eq:sni}) or (\ref{eq:sn2})  assume implicitly, something far from obvious, that singular solutions of the equation have a basin of attraction of finite volume in phase space. This implies in particular that, by linearizing the Euler-Leray equations in their form  (\ref{eq:Euler1sm}), the solution representing a singularity is stable with respect to small perturbations (we mean stable in the usual sense that small perturbations decay to zero as the "logarithmic" time $\tau=\log(t^{*}-t)$ tends to infinity as $t$ tends to $t^{*}$). Notice that contrary to the original Euler equations, the Euler-Leray equations being not reversible with respect to time $\tau$ may have stable solution in the ordinary sense, namely be attracting in all directions of phase space.  

Let us note that the above description  takes out the important question of the  space-time density  $\nu(q)$, and how  $\nu(q) $ depends on the parameters of turbulence. To answer this we need  
 to have in the theory parameters allowing to build a quantity with the physical dimension of a space-time density, namely $L^{-3} T^{-1}$, where $L$ is a length scale and $T$ a time scale. In Kolmogorov theory, turbulence in the so-called inertial domain is characterized by the power dissipated per unit mass, with the physical dimension $L^{2} T^{-3}$. No quantity with the dimension  $L^{-3} T^{-1}$ can be built by taking a power of  $L^{2} T^{-3}$, assuming that viscosity does not play a role in the formation of singular events, because they result from the inviscid dynamics only. The Kolmogorov scaling for the rate of energy dissipation by transfer of energy toward the small scales makes the quantity denoted by Kolmogorov as $\epsilon$ the only scaling parameter for turbulent velocity field with "inviscid" dissipation. This is independent of the precise mechanism of this dissipation, be it by random cascade toward smaller and smaller "vortices" continuously distributed in space or, as we shall consider it, by singular events distributed randomly in space and time. Therefore the parameter permitting to find the density of singular events in space and time cannot be linked to viscosity which is relevant near the end of the singularity formation only. This supplementary parameter must somehow depend only on the characteristics of the inviscid flow. Besides $\epsilon$, there are of course many parameters describing the turbulent fluctuations, like the mean square velocity fluctuation, or powers of the gradients of this velocity field (still partly taken into account by $\epsilon$). It could be pertinent to take as another parameter the density of turbulent kinetic energy, namely half of the mean square fluctuating velocity. This parameter is the $K$ of the $K$-$\epsilon$ theory \cite{Keps}. When seen from the point of view of the Kolmogorov-Obukhov spectrum, this parameter $K$ is also needed to make finite the total energy in the spectrum, as necessary. Therefore we shall choose as dimensionalizing parameters $K$ and $\epsilon$. It follows from this choice that a typical length and time scales of turbulence are
\begin{equation}
 \ell_{K-\epsilon} =  K^{3/2} \epsilon^{-1}  \qquad  t_{K-\epsilon} = K \epsilon^{-1}
 \textrm{.}
\label{eq:Keps}
\end{equation}
 Thanks to both scales it makes sense, for a given turbulent state to consider singularities with a  space-time density defined intrinsically as  proportional to $ \ell_{K-\epsilon} ^{-3}  t_{K-\epsilon} ^{-1}$, or
 \begin{equation}
 \tilde{\nu} \propto K^{7/2} \epsilon^{-2}
 \label{eq:nuKeps}
\end{equation} 
 without explicitly relying on the size of the turbulent channel (for instance).

At this step, because we ignore the explicit form of the singular solutions of Euler or Navier-Stokes equations, it is not possible to compute explicitly the integral on the right-hand side of (\ref{eq:sn2}). However it is possible to find how it behaves as a function of $r$, because $a_s$  depends on its arguments as prescribed by the self-similar dependence (\ref{eq:a(q)}). We set $t^{*}=0$ and leave the exponents $\alpha$ and $\beta$ undefined first in this estimate, in other words we consider a single type of  singularity associated to  a given set of exponents $\alpha,\beta$. In the collapsing region  we have the scaling $r \sim (-t)^{\beta}$,  $u_s \sim (-t)^{- \alpha}$ and $a_s \sim (-t)^{- (\alpha + 1)}$.   
Inverting the relation between $r$ and $t$ one finds 
  \begin{equation}
  t \sim r^{1/\beta} \;\;  u_s \sim r^{-\alpha/\beta}  \;\; a_s \sim r^{-(\alpha+1)/\beta}
   \textrm{.}
 \label{eq:scaling}
\end{equation}
 The  volume of the domain ${\mathcal{D}_{i}(q_{i})}$  is of order $r^{3+1/\beta}$ . Substituting those power laws in the right-hand side of equation (\ref{eq:sni}), one finds for the  contribution of singularities  with  given exponents $\alpha, \beta$,  to  $S_{n}(r)$   at short distance
  \begin{equation}
 {\mathcal{S}}_n^{(a)} (r) \sim  r^{\frac{3\beta+ 1 - n (\alpha+ 1)}{\beta}}  
 \textrm{,}
 \label{eq:P1.1}
\end{equation}
for the structure functions associated to the acceleration, and for the  structure functions associated to the velocity
  \begin{equation}
 {\mathcal{S}}_n^{(u)} (r) \sim r^{\frac{3\beta+ 1 - n \alpha}{\beta}}  
 \textrm{.}
 \label{eq:P1.v}
\end{equation}

Expression (\ref{eq:P1.1}) shows that as $n$ gets larger than $(3\beta + 1)/(\alpha +1)$,   ${\mathcal{S}}_n^{(a)} (r)$ diverges formally as $r$ tends to zero. Similarly ${\mathcal{S}}_n^{(u)} (r)$ diverges formally for $n > (3\beta + 1)/\alpha$.

With the Sedov-Taylor exponents,  ${\mathcal{S}}_n^{(a)} (r)$ diverges for $n \ge 2$, and  ${\mathcal{S}}_n^{(u)} (r)$ diverges for $n\ge 4$. 
Note first that a divergence of $S_{n}(r)$ at $r=0$ is formally inconsistent with the definition (\ref{eq:Sn}) which should be zero at $r = 0$.  The viscosity effect  is often  invoked to explain the regularization of the solution just before the divergence, however, as discussed in Sec.\ref{viscosity}, 
 with the Sedov-Taylor exponents the viscosity effect is not sufficient to dissipate the energy in the singular space-time domain and to stop the evolution to the singularity.

Besides this determination of a critical value of $n$ such that $ {\mathcal{S}}_n (r)$ changes behavior near $r = 0$, which is not very well defined from the experimental data, there is also a 
consequence of the scaling law in equation  (\ref{eq:P1.1}), namely that the exponent is an affine function of $n$, a result confirmed by the experiment, see Fig.\ref{fig:p(n)} in the next subsection.

\subsection{Experimental results}
\label{exp}
 
We use the same data as in  \cite{PC}, namely those recorded  from hot wires  in the  wind-tunnel of Modane,  by Y. Gagne et al. in the $90$s \cite{expmod}-\cite{exp2mod},  and also more recent ones obtained in 2014 in the framework of a ESWIRP European project\cite{mickael}.  In both cases we got records from a single probe. If the probe  is localized at $r_{p}$, the measured velocity at time $t$ is  
$$ {\bf{u}}_{p}(t) = u_{0} + {\bf{u}}(r_{p}-u_{0}t, t),$$
where $u_{0}$ is the mean velocity of the turbulent flow.
We assume that the Taylor approximation  of frozen turbulence is valid,  then the spatial dependence of the structure functions is replaced by a temporal dependence, with $t=r/u_{0}$. In the following the velocity fluctuation recorded by the hot wire, labelled $u$, is  assumed to be the longitudinal component of $ {\bf{u}}_{p}(t) - u_{0}$, and the Eulerian acceleration, labelled $a$, is the quantity  calculated from the increments  of this velocity, $a= (u(t_{i+1})-u(t_{i}))/t_{u}$, which approximates the partial derivative of the longitudinal velocity if  the sampling time $t_{u}$ is small enough.
The former measurements were
taken in the return vein of the Modane wind tunnel, with  Reynolds number $Re_{\lambda} =2500$,  mean velocity about $u_{0}= 20$m/s,   r.m.s. $\sigma_{u}=1.67$m/s, sampling frequency $f=1/t_{u}=25 $KHz  and   Kolmogorov time  estimated as $t_{k}=0.1$ms. In the latter case the data  were recorded behind a grid put in the test section of the wind tunnel, the Reynolds number was  about five times smaller, $Re_{\lambda}\simeq 500$, and the sampling frequency $f=250$ KHz  was ten times larger than in the ancient record.  The statistical study  based on structure functions which is described in this section stem from the ten minutes record of Y. Gagne in the $90$s, however  we have checked that recent and ancient data fully agree. 

\begin{figure}
\centerline{ 
(a)\includegraphics[height=2.in]{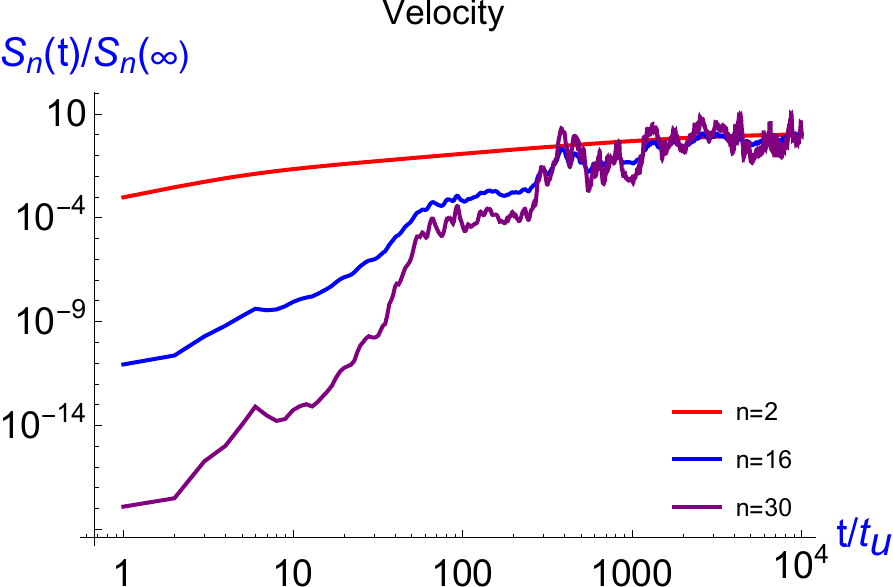}
(b)\includegraphics[height=2.0in]{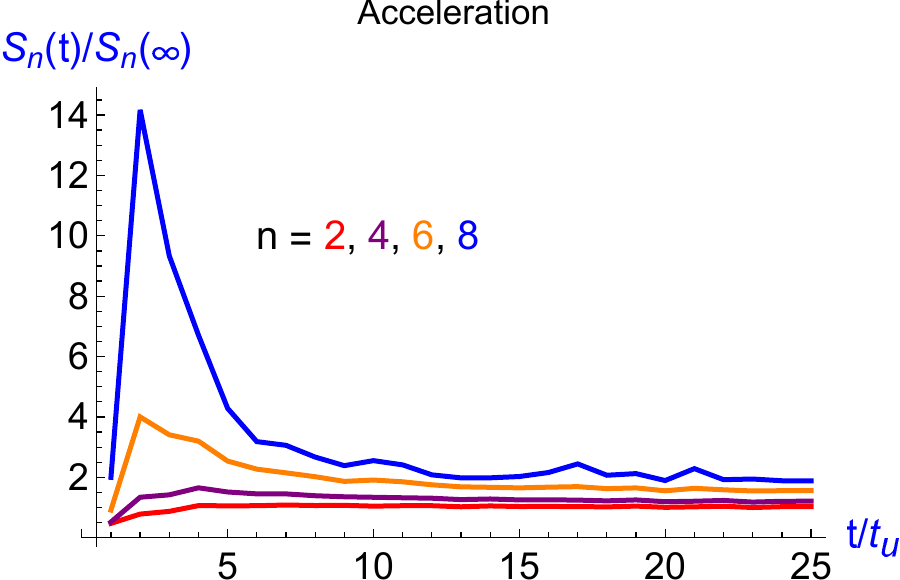}
  }
\caption{ Structure function of the velocity (a) and acceleration(b), both scaled to their value at large $r$.  }
\label{fig:Sn}
\end{figure}

We found that the structure functions for the velocity and acceleration reveal a very different behavior at large values of $n$, especially for small spatial scales (close to the Kolmogorov scale), as shown in Figs.\ref{fig:Sn}  where the  $S_{n}(t/t_{u})$ curves for the acceleration display a maximum  increasing strongly with  the order $n$, although the structure functions for the velocity grow monotonically  before they reach their constant asymptotic  value for $t $ much larger than the correlation time.  We note that the maxima of curves (b) occur practically at the Kolmogorov time ($2t_{u} \approx t_{k}$).
We also observe that $ {\mathcal{S}}_n^{(a)} (r) $
tends (rapidly) to a constant as a function of r,  as r increases.  This short distance correlation of the acceleration agrees with the hypothesis  of statistically independent  singularities made in \ref{contribution}.
 
 \begin{figure}
\centerline{ 
(a) \includegraphics[height=2.in]{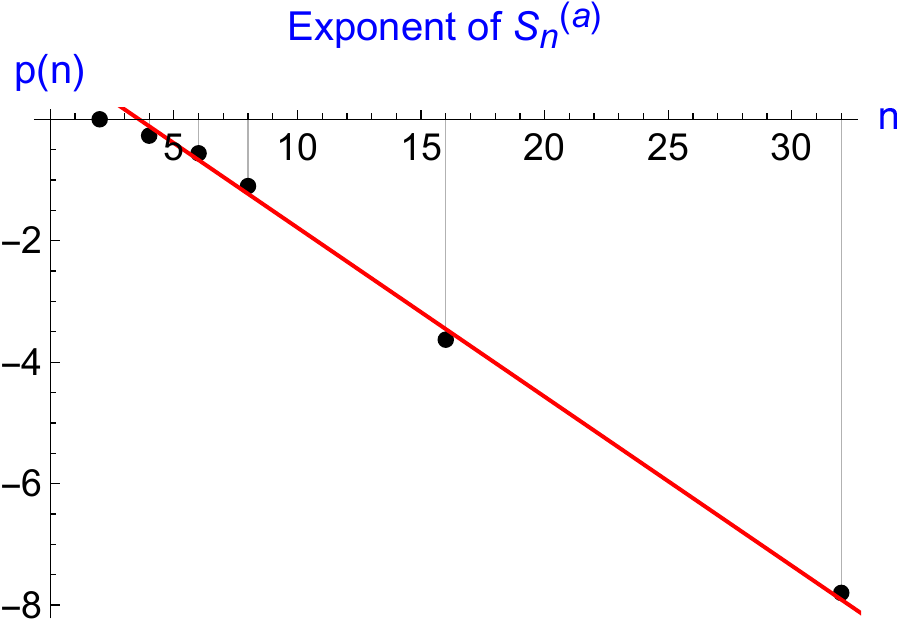}
 (b)\includegraphics[height=2.in]{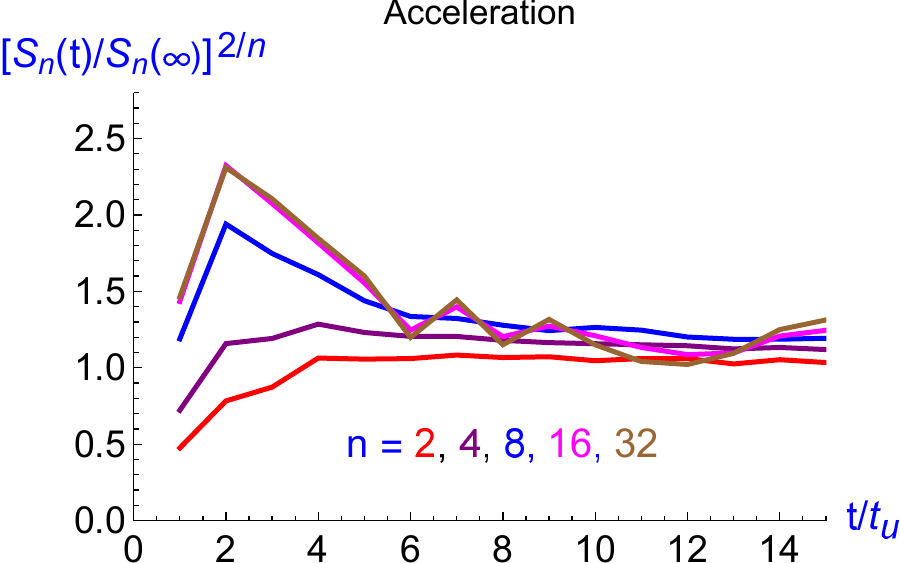}
  }
\caption{ (a) Experimental exponent $p(n)$  defined in (\ref{eq:p(n)})  with $r=u_{0}t $. The values of $p(n)$ are calculated from the  ${\mathcal{S}}_n^{(a)} (t)$ curves, in the small scale range located in the right part of the strong peak of Fig.\ref{fig:Sn}-b . (b)  Rescaled structure functions $({\mathcal{S}}_n^{(a)} (t/t_{u}))^{(2/n)} $. The quasi-linearity of  the exponent $p(n)$  at large $n$, see curve  (a),  is confirmed by curves (b):  the curves $n=16$ and $n=30$ collapse. }
\label{fig:p(n)}
\end{figure}

If Sedov-Taylor exponents were the only ones  contributing to $p(n)$, a peak should appear in structure function for acceleration {\textit{and}} velocity at large $n$. Therefore, although 
 singularities with Sedov-Taylor exponents should be present in the turbulent flow, the very different  behavior of curves (a) and (b) in Figs.\ref{fig:Sn} indicates that these solutions   have small contribution in the structure functions, that enforce the hypothesis made above, of multi-type singularities. More precisely one may deduce an effective exponent $p(n)$ from the  sharp growth of experimental curves  (b)   as $t/t_{u}$ decreases. Focusing  on the right part of the peaks in the  range  $2 \le t/t_{u}\le 10$, and setting
  \begin{equation}
 {\mathcal{S}}_n^{(a)} (r) \sim  r^{p(n)}  
 \textrm{,}
 \label{eq:p(n)}
\end{equation}
we measure an  exponent $p(n)$. We find that it  decreases linearly with $n$, as shown in Fig.\ref{fig:p(n)}, as 
  \begin{equation}
p(n)=1-0.28 n
 \textrm{.}
 \label{eq:affine-p}
\end{equation}
Comparing with the formal expression (\ref{eq:P1.1}), we conclude that, if singular events are responsible for the  striking behavior of curve (b),  they correspond in average to exponents 
  \begin{equation}
\alpha \approx -0.56   \;\; \beta\approx 1.56
 \textrm{.}
 \label{eq:alph-eff}
\end{equation}

As written in the introduction, negative values of $\alpha$ , with (\ref{eq:alphaneg}) fulfilled,  may correspond to   self-similar solutions  associated to what we called  ''weak'' singular events in the introduction, they have singular  acceleration but regular velocity.  Therefore equation ( \ref{eq:alph-eff}) open the perspective  of  the existence  of weak singularities  in the turbulent flow of Modane's  but it also deserves some remarks.

\section{Discussion}
\label{discussion}
These ''weak'' events can have several reasons to appear in the recorded signal. Those noted below deserve to be considered, beyond the fact that  all of them are able to explain that the  exponent $\alpha$ derived from the  experimental  curves, is negative.

\subsection{Role of viscosity for Sedov-Taylor singular events}
\label{viscosity}
Before focusing on possible multiplicity of $\alpha$ values, let us consider singularities with  Sedov-Taylor exponents. We are going to show that only a small part of the energy is dissipated via viscosity effects in these structures.  As generally understood viscosity enters into play  in the ultimate stage of the blowing up of the solution when $t$ approaches the formal $t^{*}$ value,  preventing the growth of the solution.  First  let us notice that  if this is true,  then viscosity is expected to act also after $t^{*}$  when the fluctuations remain large, see Fig.\ref{fig:pic-mod1}, that should make post-collapse peaks in the signal  contributing to the structure function, including at r small, in an unknown (yet) way.   Secondly, we claim that even though it is widely believed that viscosity becomes dominant as scales of turbulence become very small, this is not that obvious concerning finite time singularities with $\alpha > \beta$. This is because, as the time of blow-up approaches, the typical length scale $r$ tends to zero, but the velocity  $u$ increases as well. More precisely one may give an order of magnitude of the local Reynolds number, and then of the energy dissipated by viscosity effect. The local Reynolds number  which measures the balance between the forces of inertia and of viscosity in the fluid, is given by
  \begin{equation}
   R_e \sim  \frac{r u}{\eta} 
    \textrm{,}
 \label{eq:Re}
\end{equation}
 where $\eta$ is the kinematic viscosity of the fluid.
The velocity near the core of the singularity is of order of magnitude $u \sim (-t)^{-\alpha}$ whereas the size of the singular domain decreases like $r \sim (-t)^{\beta}$. Therefore the Reynolds number evolves as, 
  \begin{equation}
 R_e \sim  \frac{(-t)^{-\alpha +\beta} }{\eta} 
    \textrm{,}
 \label{eq:Re2}
 \end{equation}
 i.e. with a negative exponent  when $\alpha > \beta $, that occurs for instance in the Sedov-Taylor case, $\alpha = 3/5$ and  $\beta = 2/5$ which is specially attractive because  it conserves the energy.

We may  estimate  the energy dissipated by viscosity in the collapsing  space-time domain in the following way.  During the collapse,  the local energy $E$ is conserved, its order of magnitude being  $E =  \int {\mathrm{d}} r \; u^{2} \sim  r^{5} t^{-2}  $, which gives the relation  connecting the spatial scale $r$, the time scale $t$ and the energy in the singular domain,
\begin{equation}
r \sim (E t^{2})^{1/5}
  \textrm{.}
 \label{eq:Ert}
\end{equation}
Let us compare $E$ with the dissipated energy due to viscosity effect $E_{\eta}$. The dissipation rate, $ \epsilon_{\eta}=  {\mathrm{d}}E_{\eta}/ {\mathrm{d}}t$   is the  space integral 
\begin{equation}
 \epsilon_{\eta} = \eta  \int {\mathrm{d}}r  \;  ( \nabla u)^2
  \textrm{.}
 \label{eq:Eeta}
\end{equation}
Putting  (\ref{eq:Ert}) in  (\ref{eq:Eeta}) we get 
\begin{equation}
\epsilon_{\eta}  \sim \eta E^{3/5} (t^* - t)^{-4/5}
  \textrm{.}
 \label{eq:epseta}
\end{equation}
 Therefore once the equation  (\ref{eq:epseta}) for the dissipation rate $\epsilon_{\eta}$ is integrated over time $t$ until time $t^*$, it shows that only a finite energy is dissipated by viscosity, 
 \begin{equation}
E_{\eta}  << E
  \textrm{,}
 \label{eq:EetaE}
\end{equation}
all this assuming that the solution  stays close to the one of the Euler-Leray inviscid equation which is possible if the viscosity $\eta$ is small enough.  This shows that the energy dissipated by viscosity in the collapsing region is a fraction of the total available energy there,
 therefore viscosity is not obviously winning over non linear advection term and pressure near the blow-up time. This gives at least a qualitative way to build a  singular solution of Navier-Stokes once a solution of Euler-Leray has been shown to exist with the Sedov-Taylor exponents  \cite{swirl}.  Indeed if such a thing happens (singular solution with viscosity included) other physical mechanisms of regularization than viscosity should come into play, because there is no  ''physical singularity'' in a continuous medium because of the existence of atoms. One can mention two of them: at very large accelerations compressibility effects should  become relevant and some damping should be due to sound emission. 
  In dense fluids the singularity may also stop due to phenomena of higher order than viscosity in an expansion of large wavelength (or weak gradient),  named Enskog expansion. At  next order one finds formally a third order spatial derivative (although the viscosity comes at order $2$), but it happens that in dense fluids this term, named Burnett term, diverges for a non trivial reason\cite{burnett}.

\subsection{Role of dilation invariance }
\label{dilation}
The parameter $\mu(q)$ in (\ref{eq:a(q)}), related to the dilation invariance of the Euler equation, may depend on time $\tau$ or not. If it does, the evolution of this parameter
could make  post-singular  solutions  decaying with a power of $\tau=\log(t-t^{*}$), and/or in an oscillatory way, as considered in \cite{PC}, an hypothesis compatible with the burst shown in  Fig.\ref{fig:pic-mod1}.
But let consider the role of the dilation parameter in the opposite case, assuming that $\mu$ is a constant.  We set $\mu= e^{-c}$.  A self-similar solution is generally written as (\ref{eq:self-sim}).   Because of the dilation invariance of Euler (and NS) equation,  (\ref{eq:self-sim}) actually describes a family of self-similar solutions that may be also written (see Appendix) as 
    \begin{equation}
u(x,t,c)= \mu^{-1} \vert t^{*} - t \vert ^{-\alpha_{1}}  U( \mu  x \vert  t^{*} - t \vert ^{-\beta_{1}} )
\textrm{.}
\label{eq:self3}
\end{equation}
Compared to (\ref{eq:self-sim}) it appears that the exponents $\alpha,\beta$ are related to $\mu$ (or $c$) and $\tau$ via the relation
    \begin{equation}
 \alpha= \alpha_{1} +c/\tau,   \qquad   \beta= \beta_{1} -c/ \tau
 \textrm{,}
\label{eq:alf2}
\end{equation}
with 
   \begin{equation}
   c/\tau = \ln(\mu)/\ln(t^{*} -t)
    \textrm{.}
\label{eq:cstau}
\end{equation}

The expressions in (\ref{eq:self3})-(\ref{eq:alf2}) could be misleading because  one could deduce 
 that the effect of the  dilation parameter $c$ is to change the fixed exponents  $\alpha, \beta$  of the solution for $\mu=1$, into a family of time dependent exponents, that 
 is formally wrong because the solutions (\ref{eq:self-sim}) and (\ref{eq:self3}) are defined with  constant exponents $\alpha,\beta$ and $\alpha_{1}, \beta_{1}$ respectively. Nevertheless when comparing  the solution $u(x,t,c)$  with  $u(x,t,1)$, everything looks  as if the exponent  $\alpha $ is time dependent  when $\mu \ne 1$, see Appendix for more details.

We propose  an interpretation of  the discrepancy between the value of $p(n)$ in (\ref{eq:P1.1}) and the experimental one (\ref{eq:affine-p}), which is based on  the  effect of the dilation invariance of Euler  solutions on the recorded signals.  As shown in the appendix, taking into account the whole family of solutions parametrized by $\mu$ ,  a good agreement between theoretical and  experimental values of $p(n)$ is found 
 if one assumes that the main contributions to the striking peak observed in $S_{n}$ curves for large $n$ and small $r$ values, come from self-similar solutions with small amplitudes (or large $\mu$ values), namely such that
  \begin{equation}
1/\mu_{peak} \sim t^{*} -t
 \textrm{.}
 \label{eq:mup}
\end{equation}

Let us note that although this result is  different from the one  presented in \cite{PC}  the two analysis are compatible, see Appendix.  To summarize we recall that the  statistical study in \cite{PC} was performed conditionally on large acceleration values, then  it focused  on  bursts with  large amplitude. It follows that even though there are fewer events with large amplitude than  with small amplitude, the former contribute to (\ref{eq:a-u3}), although the latter contribute to $S_{n}(r) $, since there are more numerous in the flow.

\subsection{other possible causes}
Another  explanation of the discrepancy between experimental results  and the  values of $p(n)$ predicted  in  (\ref{eq:P1.1}) within the hypothesis of $\mu=1$, could be that $\mu$ depends on $\tau$. This possibility is  not considered in this section, and as shown below it requires  either that $U$ also depends on $\tau$ (not only via the dilation parameter),  or that (\ref{eq:cprime}) is fulfilled.   In that case one has to go beyond the frame of self-similar solutions. Nevertheless one can infer from (\ref{eq:alf2}) that if  a  solution  at given $\mu$ value shifts to another value of $\mu$ before or after the formal blow-up time $t^{*} $, the signal recorded by the probe should include this change of power law.

 The exponent $\alpha$  with negative values  could also belong to a continuous spectrum of possible Euler-Leray solutions.  In that case the short distance  behavior of the structure function could be dominated by the $\alpha$-negative solutions if they are more numerous and/or have high amplitudes that enhance their effect on the $S_{n}(r)$ curves.  However, let us note that the presence of such solutions goes against conservation of energy by Euler-Leray equations which conserve energy only when the exponents have Sedov-Taylor values.

 \section{Conclusion and perspectives}
 \label{Concperspect}
 This note was to  argue in favor of the occurrence of Leray-like singularities in turbulent fluids by an analysis of hot wire records. Our main point is that two  "qualitative"  features of the structure functions for the acceleration are well (if not uniquely!) explained by the existence of those singularities: first their long range behavior is  a constant function of the distance, that argue in favor of independent structures localized randomly in space-time, secondly  they exhibit a remarkable transition in their behavior at small range $r$ as the exponent $n$ increases: at $n$ "small"  there is practically no extrema in curves \ref{fig:Sn}-b, whereas a peak appears with increasing amplitude as $n$ increases. 
 Following the curves with decreasing $r$, the function tends smoothly to zero  for small $n$, whereas as $n$ gets bigger the function  is first shooting-up (right part of the peak) and ultimately decays to zero at $r = 0$ (left part). Moreover the exponent $p(n)$ in  the shooting up stage is, as predicted, an affine function of the power $n$ in the structure function. 
 
We believe that all this makes a convincing case for the existence of finite time singularities of the Euler equations. Of course this should be completed by a mathematical proof of existence of those singularities, a difficult problem. We refer the interested reader to a recent publication on this topic in the case of axisymmetric geometry with swirl\cite{swirl}. This explains how to build such a singular solution of Euler-Leray by perturbation starting from an explicit solution of the Hicks equation. 

One could also think to explain the Toms effect\cite{toms} of turbulent drag reduction in dilute solutions of polymers. No universally accepted explanation of this remarkable effect seems to exist. Supposing that dissipation in turbulent flows occurs mostly in singular events, it is reasonable to assume that long polymer molecules could stop  the evolution of a local fluctuation toward scales smaller than the size of the long polymer and so weaken the dissipation in a turbulent fluid.

\section*{Appendix: Dilation invariance  for Euler-Leray solutions }
   \label{sec: alpha1}
  The  Euler equation is invariant via a family  dilations  $\tilde{D}_{\lambda,\alpha}$ characterized by two parameters , $(\lambda,\alpha)$, because
    if $u(x,t)$ is solution, any solution   of the form    
     \begin{equation}
 ( \tilde{D}_{\lambda,\alpha} u ) (x,t)= \lambda^{-\alpha}\;  u(\lambda^{ -\beta }x, \lambda^{-1}t)    \qquad    \textrm{with} \qquad  \alpha + \beta=1,
  \label{eq:dilation}
\end{equation}
is also solution. 

We consider formal self-similar solutions of Euler equation. Close to the singularity, setting $\lambda= \vert (t^{*}-t)  \vert$ ,  
 they are generally written as 
    \begin{equation}
 u(x,t,\tau)= \vert t^{*}-t  \vert ^{-\alpha}  U(  \vert t^{*}-t  \vert^{-\beta}, \tau) 
 \textrm{,}
\label{eq:self1}
\end{equation}
 where the log-time variable 
 $$\tau=-\ln (\vert t^{*} -t \vert ),$$
  comes from a possible dependence of the reduced velocity $U$ with respect to time. This expression
does not highlight the  property associated to the invariance of solutions via a constant amplitude parameter.  This can be better seen by  taking another couple of variables,  in place of  $\lambda,\alpha$, for instance $\lambda,\mu$ 
which is  related to  $\lambda,\alpha$  by the relation $$\alpha= \alpha _{1}+  \ln (\mu) /\ln \lambda,$$ or introducing the parameter  $c = -\ln (\mu)$, it gives
  \begin{equation}
 \alpha= \alpha_{1} - c /\ln \lambda,
 \label{eq:alf1c}
\end{equation}
so that $\alpha_{1}$ is equal to $\alpha$ for $\mu=1$.
In this case  equation (\ref{eq:dilation}) is replaced by
    \begin{equation}
 ( D_{\lambda,\mu} u) (x,t)= \mu^{-1} \lambda^{-\alpha_{1}} \; u\;( \mu \lambda^{ -\beta _{1}}x, \lambda^{-1} t )  
  \textrm{.}
  \label{eq:dilation2}
  \end{equation}
and the   self-similar solutions (\ref{eq:self1})  become
    \begin{equation}
u(x,t,\mu,\tau)=\mu^{-1}  \vert t^{*}-t \vert ^{-\alpha_{1}}  U(  \mu x \vert t^{*}-t \vert ^{-\beta_{1}}, \tau)  \;\; 
\textrm{.}
\label{eq:self20}
\end{equation}

\subsubsection{Comparison between $u(x,t,\mu)$ and $u(x,t,1)$ }

We consider stationary solution of Leray-like  equation (\ref{eq:Euler1sm}), and drop the variable $\tau$ in the velocity field.
 The expression  in  (\ref{eq:self1})  is equivalent to (\ref{eq:self20})  with $\alpha$ linked to the couple ($\alpha_{1},\mu$) by the relation (\ref{eq:alf2}).  As written in the text,
any self-similar solution $u(x,t,\mu)$ which satisfies (\ref{eq:self20}) , with a given $\mu$,  has an amplitude proportional to  a fixed coefficient $1/\mu$ which increases with respect to time as $ \vert (t^{*}-t) \vert ^{-\alpha_{1}} $, with constant $\alpha_{1}$  value during the pre (or post)-collapse. However  expressions (\ref{eq:self3})-(\ref{eq:alf2}) mean that for a given value of  time $ t < t^{*}$,  and given $\alpha_{1}$  (the asymptotic value at singular time) the stretching of the solution $u(x,t,\mu)$ is different from the stretching of $u(x,t,1)$, except at singular time $t=t^{*}$. More precisely  one can say 

- either that $u(x,t,\mu)$ at time $t$ is the same as  $u(x,t',1)$ at time $t'$ such that $t^{*}-t'= (t^{*} -t)^{1+c/ \alpha_{1} \tau}$,

- or else that $u(x,t,\mu)$ is stretched at time $t$ as  the solution $u(x,t,1)$  (at the same time) associated to the exponent $\alpha$ given by (\ref{eq:alf2}).

\begin{figure}
\centerline{ 
\includegraphics[height=2.in]{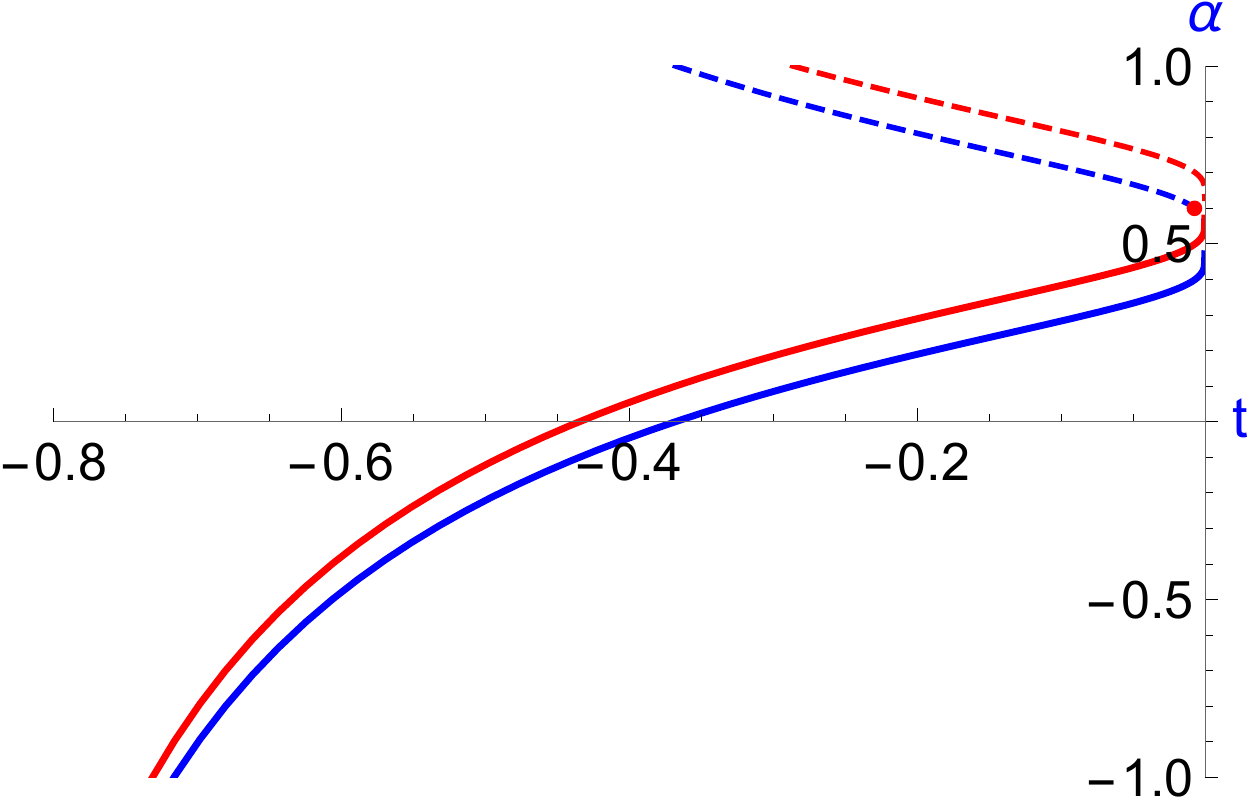}
  } 
\caption{  Exponent $ \alpha$  of a self-similar solution with dilation parameter $c=\pm 0.5$, or $\mu=0.6$ for the dashed curves, $\mu=1.65$ for the solid curves,
versus $t $  before the singular time $t^{*}=0$ , for $\alpha_{1}=1/2$ (blue curve), and $\alpha_{1}=3/5$ (red curve) .   
The  value  of $\alpha$ at singular time, doesn't depend  on $c$,  it only depends on   $ \alpha_{1} $. The red point  on the blue curve is the  Sedov-Taylor  value for $\alpha_{1}$.}
\label{fig:alf}
\end{figure}

 Fig. \ref{fig:alf} displays the evolution of $\alpha$ versus $ t^{*} -t$ for $\mu=e^{-0.5}$ (dashed lines) and  $\mu=e^{0.5}$ (solid lines)  for two values of the asymptotic exponent $\alpha_{1}$, the Sedov-Taylor value and the NS  value.   The solid curves are for  $\mu $ larger than unity, they correspond to a solution $u(x,t,\mu)$ with  amplitude and width smaller than $u(x,t,1)$. 
 They are interpreted  as responsible for the peak observed in $S_{n}^{(a)}$, see below. 
  Dashed curves are for $\mu$ smaller than unity,  namely for the dilated solution $u(x,t,\mu) $  with amplitude and width larger than u(x,t,1).

\subsubsection{Link with experiments}
We propose an interpretation of  this discrepancy between the value of $p(n)$ in (\ref{eq:P1.1}) and the experimental one (\ref{eq:affine-p}) by looking at the role of  the  dilation invariance of Euler  (and NS)  solutions.  

Assuming, as done in this section, that a  family of  Euler-Leray solutions exist with a continuum of $\mu$ values, the signal recorded at a given place $x$ must reflect a large variety of exponent values $p(n,\mu)$, or equivalently $p(n,c)$. More precisely, for a given $\mu$ value,  one expects that the behavior of the structures functions for the acceleration is given by (\ref{eq:P1.1}) with  $\alpha$ given by (\ref{eq:alf2}), so that in average, one get
  \begin{equation}
 {\mathcal{S}}_n^{(a)} (r) \sim < r^{p(n,c)}>_{c} 
 \textrm{,}
 \label{eq:Smu}
\end{equation}
where $<>_{c}$ means an average over $c$ values, $p(n,c)$ is  the exponent in  (\ref{eq:P1.1})  with $\alpha$ replaced by the  expression  (\ref{eq:alf2}) , that gives 
  \begin{equation}
  p(n,c) = \frac{ 3(\beta_{1}  - c/\tau) + 1}{\beta_{1}  - c/\tau} - n \frac{\alpha_{1} +c/\tau +1}{\beta_{1}  - c/\tau}.
 \label{eq:pnc}
\end{equation}
A rough estimate of the average (\ref{eq:Smu}) can be given  by assuming that the values of $\mu$ contributing to this expression are those around a given $\mu_{p}$ which is to be deduced from the experimental results. To do that we identify the slope $-\frac{\alpha_{1} +c/\tau +1}{\beta_{1}  - c/\tau}$ of the exponent $p$ in (\ref{eq:pnc})  with the experimental value $-0.28$. We get  the relation $1/\mu \approx (t^{*} -t) ^{1.16}$. This result  allows to conclude that if the main contributions to the striking peak observed in $S_{n}$ curves for large $n$ and small $r$ values come from Leray singularities, they are due  to the small amplitudes  solutions with
  \begin{equation}
1/\mu_{p} \sim t^{*} -t
 \textrm{.}
 \label{eq:mup}
\end{equation}

Let us now return to our previous analysis of Modane's data presented in \cite{PC} . There we made a statistical studies of  the coupling between  large acceleration and velocities values recorded at the same time, namely we didn't look at the spatial autocorrelation  of the acceleration contrary to what we do here for the study of  $S_{n}(r) $ (with the same  data).  
In \cite{PC} we got a very good agreement between the scaling predicted  $\gamma \sim u^{8/3}$ (with Sedov-Taylor-singularities) and those deduced by the record. We deduce that  this study focused on  the range of self-similar solutions with small values of $c/\tau$ , or amplitude of order unity, 
  \begin{equation}
1/\mu  \sim 1
 \textrm{,}
 \label{eq:mupc}
\end{equation}
In summary the role of the dilation invariance could amount to select bursts of different  amplitude  according to the  statistical   analysis performed,
Sedov-Taylor exponents  are detected from large amplitude bursts,  larger than those  which contribute to $S_{n}(r) $.

Another  explanation of the discrepancy between experimental results  and the  values of $p(n)$ predicted  in  (\ref{eq:P1.1}) within the hypothesis of $\mu=1$, could be that $\mu$ depends on $\tau$. This possibility is  not considered in this section, and as shown below it requires  either that $U$ also depends on $\tau$ (not only via the dilation parameter),  or that (\ref{eq:cprime}) is fulfilled.   In that case one has to go beyond the frame of self-similar solutions, nevertheless one can infer from (\ref{eq:alf2}) that if the a  solution  at given $\mu$ value shifts to another value of $\mu$  before or after the formal blow-up time $t^{*} $, the signal recorded by the probe should include this change of power law.

\subsubsection{Stationary solution of  time dependent  Leray equation (\ref{eq:Euler1sm})}

The function $U$ depends on the  reduced spatial variable   $X=x \vert t \vert ^{-\beta}$, moreover it may depend on $\tau$ or not.  If $U$ depends on $\tau$,
the Euler equation is equivalent to  what we  call  the ''time dependent  Leray''  equation (\ref{eq:Euler1sm}) for $U(X,\tau)$, which can also be written as
   \begin{equation}
   \partial_{\tau}U +  (\alpha \tau +c)'  U + (\beta \tau -c)' (X \cdot \nabla)U + (U\cdot \nabla)U +\nabla W =0   
\textrm{,}
\label{eq:Leray1}
\end{equation}
where  the prime exponent is for the time derivative  $\partial_{\tau}='$.
 If $c$  (or $\mu$) doesn't depend on $\tau$, (\ref{eq:Leray1}) doesn't depend on $c$.  But if one assumes the opposite, namely that $\alpha$ and $\mu$  (or $c$) depend on $\tau$, then a stationary solution  of (\ref{eq:Leray1}) exists only if 
   \begin{equation}
( \alpha \tau +c )' =0
 \textrm{.}
\label{eq:cprime}
\end{equation}

\section*{Acknowledgments}
 The authors are very grateful to Jean Ginibre for very useful discussions.

{}

\thebibliography{99}
\bibitem{Euler} L. Euler, "Principes g\'en\'eraux du mouvement des fluides", M\'emoires de l'Acad\'emie de Berlin (1757). 
 \bibitem{leray} J. Leray, "Essai sur le mouvement d'un fluide visqueux emplissant l'espace", Acta Math.  {\bf{63}} (1934) p. 193 - 248. 
   \bibitem{PC} Y. Pomeau, M. Le Berre and T. Lehner,  C.R. M\'ec. Paris, {\bf{347}} (2019) p. 342. in special  issue  dedicated to Pierre Coullet.; ArXiv:1806.04893v2. 
    \bibitem{Keps}  B.E. Launder, D.B. ; Spalding, "The numerical computation of turbulent flows". Computer Methods in Applied Mechanics and Engineering, {\bf{3}}, (1974)  p. 269- 289.
     \bibitem{expmod}   Y. Gagne, 'Etude exp\'erimentale de l'intermittence et des singularit\'es dans le plan complexe en turbulence d\'evelopp\'ee, PhD thesis Universit\'e de Grenoble 1 (1987).
 \bibitem{exp2mod}  H. Kahalerras, Y. Mal\'ecot, Y. Gagne, and B. Castaing, Intermittency and Reynolds number,
 Phys. of Fluids 10 (1998) 910.  
   \bibitem{mickael}    M. Bourgoin, C. Baudet, N. Mordant, T. Vandenberghe et al. ,   Investigation of the small-scale statistics of turbulence in the Modane S1MA wind tunnel , CEAS Aeronaut J., published online, July 2017. DOI 10.1007/s13272-017-0254-3  
 \bibitem{YP} Y. Pomeau, "Singularit\'e dans l'\'evolution du fluide parfait", C. R. Acad. Sci. Paris  {\bf{321}} (1995), p. 407 -411
   \bibitem{CJ} C. Josserand, Y. Pomeau and S. Rica, Finite-time localized singularities as a mechanism for turbulent dissipation, in preparation.
     \bibitem{swirl} Y. Pomeau, M. Le Berre, Blowing-up solutions of the axisymmetric Euler equations for an incompressible fluid, arXiv:1901.09426. 
     \bibitem{burnett} S. Chapman and T. G. Cowling, The Mathematical Theory of Non-Uniform Gases (Cambridge University Press, Cambridge, UK, 1970), 3rd ed.
;  P. R\'esibois and M. de Leener, Classical Kinetic Theory of Fluids (John Wiley and Sons, New York, 1977). Notice however that, because of the long tails of time correlations of dense fluids, the Burnett coefficients do not exist in dense fluids similarly as the usual transport coefficients do not exist in 2D dense fluids. Burnett transport coefficients exist in the low density limit only when Boltzmann equation applies. 
     \bibitem{toms} B. A. Toms "Observation on the flow of linear polymer solutions through straight tubes at large Reynolds numbers" Proceedings of the First International Congress of rheology Amsterdam, Vol. II, p.135 - 141 (North Holland 1949).
    
 \endthebibliography
\end{document}